\documentclass[conference]{IEEEtran}
\IEEEoverridecommandlockouts
\usepackage{amsmath,amsfonts}
\usepackage{algorithmic}
\usepackage{algorithm}
\usepackage{array}
\usepackage[caption=false,font=normalsize,labelfont=sf,textfont=sf]{subfig}
\usepackage{textcomp}
\usepackage{stfloats}
\usepackage{url}
\usepackage{verbatim}
\usepackage{graphicx}
\usepackage{cite}
\usepackage{csquotes}
\usepackage{amssymb}
\usepackage{nopageno}

\usepackage{tikz,stackengine,mathtools}
\usetikzlibrary{matrix,calc}
\newsavebox{\foobox}

\usepackage{tikz}
\usetikzlibrary{arrows,shapes.misc,chains,scopes}
\usepackage{pgfplotstable}
\pgfplotsset{compat=1.15}

\def\C{\mathcal{C}}

\def\L{\mathcal{L}}

\def\R{\mathbb{R}}

\def\Xhat{ \widehat{x} }

\def\lAPP{ \lambda^{\text{APP}} }
\def\lEXT{ \lambda^{\text{ext}} }

\def\lGRANDAPP{ \lambda^{\text{SOGRAND-APP}} }

\title{SOGRAND decoding of LDPC codes}
\begin{document}

\author{\IEEEauthorblockN{Ken R. Duffy and Jiewei Feng}
\IEEEauthorblockA{\textit{Dept. of Mathematics \& Dept. of ECE} \\
\textit{Northeastern University}\\
Boston, USA \\
\{k.duffy,feng.ji\}@northeastern.edu}
\and
\IEEEauthorblockN{Lukas Rapp and Muriel M{\'e}dard}
\IEEEauthorblockA{\textit{Research Laboratory for Electronics} \\
\textit{Massachusetts Institute of Technology}\\
Cambridge, USA \\
\{rappl,medard\}@mit.edu}
\thanks{This material is based upon work supported by the Defense Advanced Research Projects Agency (DARPA) under Contract No. HR00112120008.} 
}

\maketitle

\begin{abstract}
Long forward error correction codes are typically constructed by concatenating shorter component codes that are then decoded through iterative Soft-Input Soft-Output (SISO) of their components. The recently introduced Soft Output Guessing Random Additive Noise Decoding (SOGRAND) has been shown to enable accurate SISO component decoding for a broad range of component codes. Here we establish that by specializing its SISO computation to Single Parity Check codes, SOGRAND offers an alternative existing Check Node (CN) update for decoding Low Density Parity Check codes. Simulation results demonstrate similar or better decoding performance than Gallager's sum-product algorithm and norm-min-sum, while offering two distinct low complexity, hardware friendly CN update algorithms.
\end{abstract}

\begin{IEEEkeywords}
LDPC Codes, Soft Input, Iterative Decoding, GRAND, SOGRAND
\end{IEEEkeywords}

\thispagestyle{empty}

\section{Introduction}
The standard mechanism for constructing long, powerful error correction codes is to concatenate short component codes. In the presence of soft input, such constructions can be efficiently decoded by iterative means using a SISO decoder for each component code in conjunction with belief propagation, e.g. \cite{costello2007channel,richardson2008modern,moon2020error}. Designs of this sort include turbo codes \cite{berrou_1993_turbo} as used in 4G, 
low-density parity-check (LDPC) codes \cite{gallagher1962_lowdensity, mackay1997near, richardson2001capacity,  richardson2001efficient, mackay2003information,mansour2003high,richardson2018design} as used in 5G, turbo product codes \cite{pyndiah_1998,moon2020error,rowshan2024channel} as used in WiMAX, staircase codes \cite{smith2012staircase} and the OFEC code, etc.

A central element to performance is the accuracy of the SISO component code decoder and its ease of implementation in hardware. Low density parity check (LDPC) codes are constructed with single parity check (SPC) component codes. Perfect soft output (SO) can be computed from an SPC code using Gallager's ingenious sum-product algorithm (SPA) \cite{gallagher1962_lowdensity}. In practical implementations, approximations to the SPA, such as normalized min-sum (NMS), are employed as they CN enable lower complexity implementation in hardware with similar decoding performance, e.g. \cite{mansour2003high,zhang2010efficient,hailes2015survey,lee2022multi,jang2024area,renAGeneralized2024}.  

Revisiting SOGRAND \cite{yuanSOGRAND25} and specializing its application to SPC codes, here we introduce two novel CN update rules. In doing so, we establish new means to get the same or better performance as the SPA-based approaches through a CN update algorithm that is based on a distinctive premise that lends itself to implementation in hardware. The key philosophical difference is that while SPA, and the approximations that follow from it, focus directly on evaluating a marginal per-bit extrinsic log-likelihood ratio (LLR) updates, SOGRAND instead identifies block-wise beliefs that can be marginalized to form per-bit extrinsic LLR updates.

The rest of this paper is organized as follows. Section \ref{sec:SPA} briefly reviews SPA and its NMS approximation. Section \ref{sec:SOGRAND} introduces SOGRAND and its specialization to SPC codes. Section \ref{sec:LDPC SOGRAND} introduces two new SOGRAND CN update algorithms. Section \ref{sec:perfeval} provides a performance evaluation for 5G LDPC codes. Section \ref{sec:discussion} ends with a discussion.

\section{SPA and its approximations}
\label{sec:SPA}
Given a vector of log-likelihood ratios, $\lambda^n\in\R^n$, 
Gallager's well-known SPA for an SPC establishes that the a posteriori LLR of bit $i$ exactly equals $\lAPP_i = \lambda_i+\lEXT_i$, where $\lEXT_i$ is the extrinsic information, which can be calculated as
\begin{align*}
\lEXT_i = 2\tanh^{-1}\left(\prod_{j\neq i}\tanh\left(\frac{\lambda_j}{2}\right)\right).
\end{align*}
For each $i$, to evaluate this precisely requires the computation of relatively involved functions, $\tanh$ and $\tanh^{-1}$, though both can be approximated using, for example, lookup tables and clever interpolation or related approaches \cite{mansour2003high}. As importantly, for each $i\in\{1,\ldots,n\}$ the computation is distinct as it requires a ``leave one out'' approach to the inner argument.

Leveraging the fact that $-\ln\tanh(x/2)$ is a steep function of $x$, the min-sum approximation side-steps the computation of $\tanh$ and $\tanh^{-1}$, by observing 
\begin{align*}
\lEXT_i \approx \left(\prod_{j\neq i}\text{sign}(\lambda_j)\right)  \min_{j\neq i}|\lambda_j|.
\end{align*}
The NMS approach improves the performance of that approximation by damping the extrinsic information by a factor
\begin{align*}
\lEXT_i \approx \alpha\left(\prod_{j\neq i}\text{sign}(\lambda_j)\right) \min_{j\neq i}|\lambda_j|,
\end{align*}
where $\alpha\in(0.75,0.8)$ is typically chosen to get similar, or sometimes better, performance than SPA. While NMS completely avoids evaluating $\tanh$ and $\tanh^{-1}$, and with a suitable factor $\alpha$ can get similar performance to the SPA, to evaluate it requires the identification of a minimum of each ``leave one out'' collection $\{j\neq i\}$ for each $i\in\{1,\ldots,n\}$. Efficient circuits for those evaluations are what hardware designers seek to achieve. Here, we will compare the new SOGRAND CN updates with both SPA and NMS.  

\section{SOGRAND}
\label{sec:SOGRAND}
Guessing Random Additive Noise Decoding (GRAND) is a family of code-agnostic decoding algorithms that can accurately decode any moderate redundancy code \cite{duffy_capacity-achieving_2019, duffy2026guessing}. GRAND algorithms function by sequentially inverting putative noise effects, ordered from approximately most to least likely according to channel properties and soft information, from received signals. The first codeword yielded by inversion of a noise effect is a near maximum-likelihood decoding. As this procedure does not depend on codebook structure, GRAND can decode any moderate redundancy code. 
Both hard and soft input GRAND variants have been established, e.g. \cite{solomon20,duffy2022_ordered,abbas2021list,An22,chatzigeorgiou2023symbol,Duffy23ORBGRANDAI,chatzigeorgiou2024guessing,abbas2025improved,rowshan2025segmented}, with efficient hardware implementations, e.g. \cite{riaz2021multicodegrand,blanc2024GRANDABDecoder,Riaz25ORBGRAND,Kizilates25ORBRGANDAI}, and syntheses, e.g. \cite{abbas2021orbgrand, condo2021fixed, chu2023efficient,ji2025efficient} demonstrating the accuracy, flexibility and energy efficiency of GRAND decoding strategies.

SOGRAND has recently established that any soft input GRAND algorithm can readily provide accurate block-wise SO, even with a single decoding \cite{yuanSOGRAND25}. That is, each decoding can be accompanied by an accurate estimate of the likelihood that the decoding is correct. When compared to the classic list decoding block-wise SO approximation of Forney \cite{forney1968_exponential}, implicitly underlying Pyndiah's bit-wise SO \cite{pyndiah_1998}, the key advance is the provision of an estimate that the correct decoding is not an element of the list, even for a list of size one. By considering Guessing Codeword Decoding (GCD) \cite{ma2024guessing} and Successive Cancellation decoding of Polar-like codes through the lens of GRAND it has been established that they too can generate accurate SO \cite{duffy2025soft,yuan2025soft}, as can other designs, e.g. \cite{Janz25soft}. A taped out chip has established that GRAND SO can be computed without increasing the latency of decoding \cite{Kizilates26SOGRAND}.

A detailed derivation of SOGRAND's SO formula can be found in \cite{yuanSOGRAND25}. Here, we solely recapitulate the algorithmic result.  Assume that $k$ information bits are encoded into a codeword of length $n$ bits, $x^n\in\{0,1\}^n$, that is modulated and transmitted. Given a vector of log-likelihood ratios, $\lambda^n\in\R^n$, which results in the hard decision sequence $y^n= 1_{\lambda_i < 0}$, 
we have that $y^n = x^n \oplus N^n$,
where addition, $\oplus$, is in the binary field and $N^n$ indicates which bits have been flipped. 

A GRAND decoder queries the noise effect sequences $\{z^{n,i}:i=1,2,\ldots\}$ and asking if $y^n\oplus z^{n,i}$ is in the codebook. For $L\geq1$, letting $q_1<q_2<\cdots<q_L$ denote the query numbers at which codebook elements are identified, $\Xhat^{n,q_l}=y^n\oplus z^{n,q_l}$, 
then, if codebook structure doesn't influence the query order, the blockwise soft output is the a posteriori probability that $\Xhat^{n,q_l}$ is the transmitted codeword is $p_{X^n|\Lambda^n}(\Xhat^{n,q_l}|\lambda^n) \approx$
\begin{align}
    \frac{p_{N^n|\Lambda^n}( z^{n,q_l}|\lambda^n)}{\displaystyle \sum_{i=1}^L p_{N^n|\Lambda^n}(z^{n,q_i}|\lambda^n)+ \left(1-\sum_{i=1}^{q_L} p_{N^n|\Lambda^n}(z^{n,i}|\lambda^n)\right) 2^{k-n}}.
    \label{eq:GRANDAPP}
\end{align}

From eq. \eqref{eq:GRANDAPP}, it can be seen that the estimated likelihood that the correct decoding is not in the list is one minus the sum of the values in eq. \eqref{eq:GRANDAPP},
\begin{align*}
1-\sum_{l=1}^Lp_{X^n|\Lambda^n}(\Xhat^{n,q_l}|\lambda^n).   
\end{align*}

A binary code is called even if every codebook element has zero parity. For such a code, the parity of the demodulated sequence and the noise effect must be the same,
\begin{align*}
    \rho = \bigoplus_{i=1}^n y_i = \bigoplus_{i=1}^n \left(x_i+N_i\right) =  \bigoplus_{i=1}^n N_i \in\{0,1\}.
\end{align*}
If $\rho=0$ then an even number of bit flips must have occurred and the Hamming weight of $N_i$ is necessarily  even, while if $\rho=1$ the Hamming weight of $N_i$ must be odd. As one of the core soft input query generators, the ORBGRAND generator \cite{duffy2022_ordered}, fractionates queries by Hamming weight, knowledge of the parity of demodulated sequence can be exploited to avoid producing queries that would necessarily not identify a codeword \cite{rowshan2024channel}. 

If an even code is used but that query-avoiding feature is not employed, then eq. \eqref{eq:GRANDAPP} remains accurate. If, however, the even code property is used to inform query order, then a correction to the SO is necessary \cite{feng2025SO}. In particular, by the well-known formula due to Gallager, the likelihood that the noise is even is
\begin{align}
     \phi = p_{\bigoplus_{i=1}^n N_i|\Lambda^n}\left(0|\lambda^n\right) = 
     \frac12 \left(1+\prod_{i=1}^n\left(1-\frac{2e^{-|\lambda_i|}}{1 + e^{-|\lambda_i|}}\right) \right),
     \label{eq:phi}
\end{align}
and the $1$ in the denominator of eq. \eqref{eq:GRANDAPP} should be replaced with 
\begin{align}
    \Psi = 
    \begin{cases}
        \phi & \text{ if } \rho=0,\\
        1-\phi & \text{ if } \rho=1,
    \end{cases}
    \label{eq:Psi}
\end{align}
while the $n$ replaced with $n$-1. This gives $p_{X^n|\Lambda^n}(\Xhat^{n,q_l}|\lambda^n)\approx$
\begin{align}
    \frac{p_{N^n|\Lambda^n}( z^{n,q_l}|\lambda^n)}{\displaystyle \sum_{i=1}^L p_{N^n|\Lambda^n}( z^{n,q_i}|\lambda^n)+ \left(\Psi-\sum_{i=1}^{q_L} p_{N^n|\Lambda^n}(z^{n,i}|\lambda^n)\right) 2^{k-n+1}}.
    \label{eq:GRANDAPPeven}
\end{align}

To evaluate eq. \eqref{eq:GRANDAPP} or \eqref{eq:GRANDAPPeven} requires the computation of each $p_{N^n|\Lambda^n}( z^{n,i}|\lambda^n)$ along with its running sum, storing $p_{N^n|\Lambda^n}( z^{n,q_l}|\lambda^n)$ for each codeword that is identified, as well as $\Psi$ in \eqref{eq:phi}. In the taped out SOGRAND chip, 
multiple patterns are generated per clock-cyle and tested for codebook membership. The likelihoods $p_{N^n|\Lambda^n}( z^{n,i}|\lambda^n)$ are computed in same clock cycle, hence the SO introduces no additional latency \cite{Kizilates26SOGRAND}.

\section{LDPC SOGRAND CN Update Rules}
\label{sec:LDPC SOGRAND}
While the approach in Section \ref{sec:SPA} focuses on each individual per-bit $\lAPP_i$, a codebook-centric view suggests an alternative formulation. First note that a SPC $(k+1,k)$ code is an even code for any $k$. Moreover, with an SPC, a noise effect $z^n$ identifies a codeword, i.e. $y^n\oplus z^n$ is a codeword, if and only if $z^n$ has parity $\rho$. As a result, we have this code-book centric representation:
\begin{align*}
    \lAPP_i 
    &
    =
    \log\left(
    \frac{ \displaystyle \sum_{z^n:y_i\oplus z_i=0,\oplus z_i=\rho}p_{N^n|\Lambda^n}(z^n|\lambda^n)}{ \displaystyle \sum_{z^n:y_i\oplus z_i=1,\oplus z_i=\rho}p_{N^n|\Lambda^n}(z^n|\lambda^n)}
    \right).
\end{align*}
The numerator is the likelihood of all valid codewords given the demodulated signal that have $0$ in bit position $i$ while the denominator is the likelihood of all valid codewords that have a $1$ in that position. Complete evaluation of the likelihood of all $2^k$ noise effect sequences would evaluate the same quantity as SPA does. SOGRAND proposes an alternative approximation based on a small list of codewords.

Without using the even code property, if a collection of noise effect queries, $\C = \{z^{n,j}:j=1,\ldots\}$, are made such, with
$\L=\{z^n\in\C:\oplus z_i=\rho\}$ being those of correct parity that identify codewords, then  $\lGRANDAPP_i \approx$
\begin{align}
    \log\left(
    \frac{ \displaystyle \sum_{z^n\in\L:y_i\oplus z_i=0 }2p_{N^n|\Lambda^n}(z^n|\lambda^n) + (1-\eta) \frac{e^{\lambda_i}}{1+e^{\lambda_i}}
    }{ \displaystyle \sum_{z^n\in\L:y_i\oplus z_i=1} 2p_{N^n|\Lambda^n}(z^n|\lambda^n) + (1-\eta) \frac{1}{1+e^{\lambda_i}}}
    \right),
    \label{eq:LDPC-SOGRANDU}
\end{align}
where $\eta = \sum_{z^n\in\C} p_{N^n|\Lambda^n}(z^n|\lambda^n)$ is the sum of the probability of all queried sequences. This provides one SOGRAND CN update rule from which the extrinsic information can be calculated as $\lEXT_i \approx \alpha (\lGRANDAPP_i-\lambda_i)$.

For the second SOGRAND rule, if only queries of the correct parity are made, then every query would necessarily identify a codeword without any further check, $k-n+1=0$, and the even code SO GRAND update in eq. \eqref{eq:GRANDAPPeven} simplifies to $p_{X^n|\Lambda^n}(\Xhat^{n,q_l}|\lambda^n)\approx p_{N^n|\Lambda^n}( z^{n,q_l}|\lambda^n)/\Psi.$ As every noise effect of the correct parity identifies a codeword, it is possible to make a fixed number of queries, $L$, and guarantee $L$ codewords are found. Define $\L=\{z^n\in\C:\oplus z_i=\rho\}$, as above, and 
then $\lGRANDAPP_i \approx$
\begin{align}
    \log\left(
    \frac{ \displaystyle \sum_{z^n\in\L:y_i\oplus z_i=0}p_{N^n|\Lambda^n}(z^n|\lambda^n) + (\Psi-\eta) \frac{e^{\lambda_i}}{1+e^{\lambda_i}}
    }{ \displaystyle \sum_{z^n\in\L:y_i\oplus z_i=1}p_{N^n|\Lambda^n}(z^n|\lambda^n) + (\Psi-\eta) \frac{1}{1+e^{\lambda_i}}}
    \right),
    \label{eq:LDPC-SOGRAND}
\end{align}
from which the extrinsic information can be calculated, $\lEXT_i \approx \alpha (\lGRANDAPP_i-\lambda_i)$.

Once a list of $\L$ noise effects is provided, to evaluate the SOGRAND LDPC CN update in eq. \eqref{eq:LDPC-SOGRANDU} requires computation of
$p_{N^n|\Lambda^n}(z^n|\lambda^n)$ for each $z^n\in\L$ as well as their sum.
If the update in eq. \eqref{eq:LDPC-SOGRAND} is used, $1/2$ as many sequences can be used, but $\Psi$ in eq. \eqref{eq:Psi} must be calculated.
For the LDPC in 5G NR standard, to match SPA performance we shall find $L$ of order $8$ to $10$ suffices, where $|\C|=2L$ for the first update rule. As a result, existing hardware demonstrates that all these quantities can be calculated in parallel in a single clock cycle in a small, efficient circuit \cite{Kizilates26SOGRAND}. All that remains is the question of how the queries should be determined. 

\begin{figure}[h]
\begin{center}
\includegraphics[width=1\columnwidth]{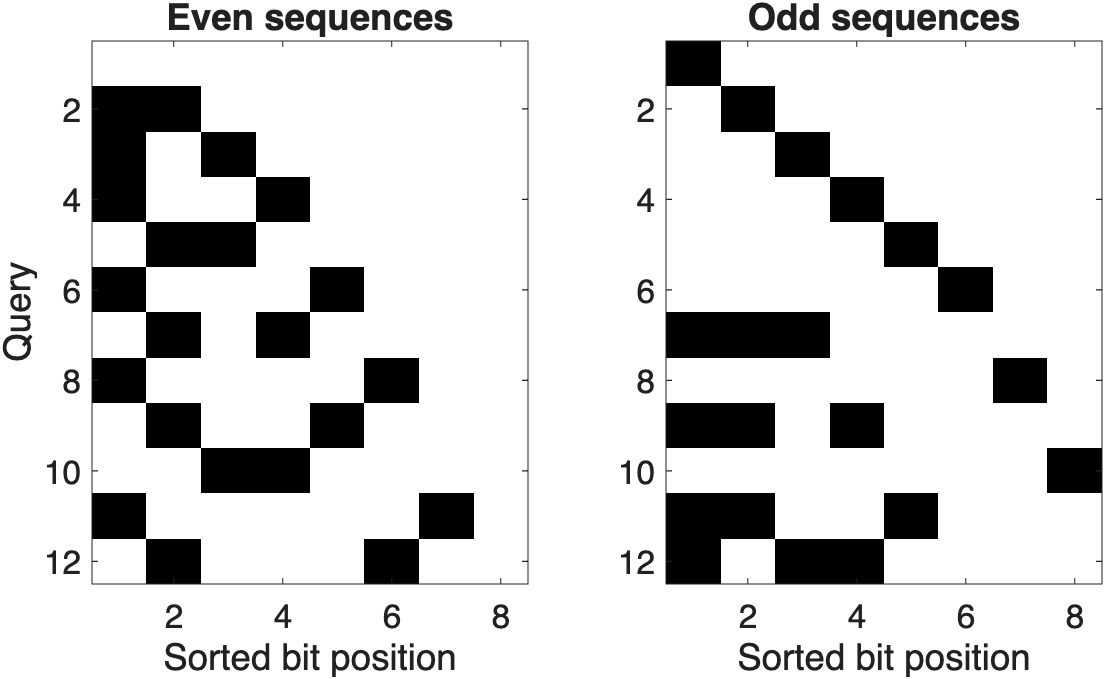}
\end{center}
\caption{If zero intercept ORBGRAND is used to create $\C$ with $L=24$ noise effect sequences, the left hand side shows even queries
and the right odd queries, one which would form $\L$ depending on $\rho$. Each row corresponds to a query. With reliabilities rank ordered from least to greatest, a black square indicates that
a bit should be flipped. Only the $7$ least reliable bits will be flipped for the even queries and the $8$ least reliable for the odd.} 
\label{fig:SOGRAND seqs}
\end{figure}

While there are other possibilities, here we use a simple fixed table that defines $2L$ patterns, $L$ even and $L$ odd, for the SOGRAND CN update based on ORBGRAND \cite{duffy2022_ordered}. In this design, given input LLRs, $\lambda^n$, for each input it suffices to identify a fixed number of least reliable bits in rank order. 

For $L=12$, for example, Fig. \ref{fig:SOGRAND seqs} plots these $2L$ sequences, with the left hand plot being the even sequences and the right hand plot being the odd sequences. To use these tables, for each received sequence it is sufficient to identify no more than the $8$ least reliable bits, which can be achieved in a modest sized circuit in order 2 clock cycles, and then reference the table to determine which bits are to flipped. As such, with $L=12$ and this ORBGRAND table, based on existing hardware designs, the complete SOGRAND CN update with either rule should take 3 clock cycles or less. 

\section{Results}
\label{sec:perfeval}

\begin{figure}[h]
\begin{center}
\includegraphics[width=1\columnwidth]{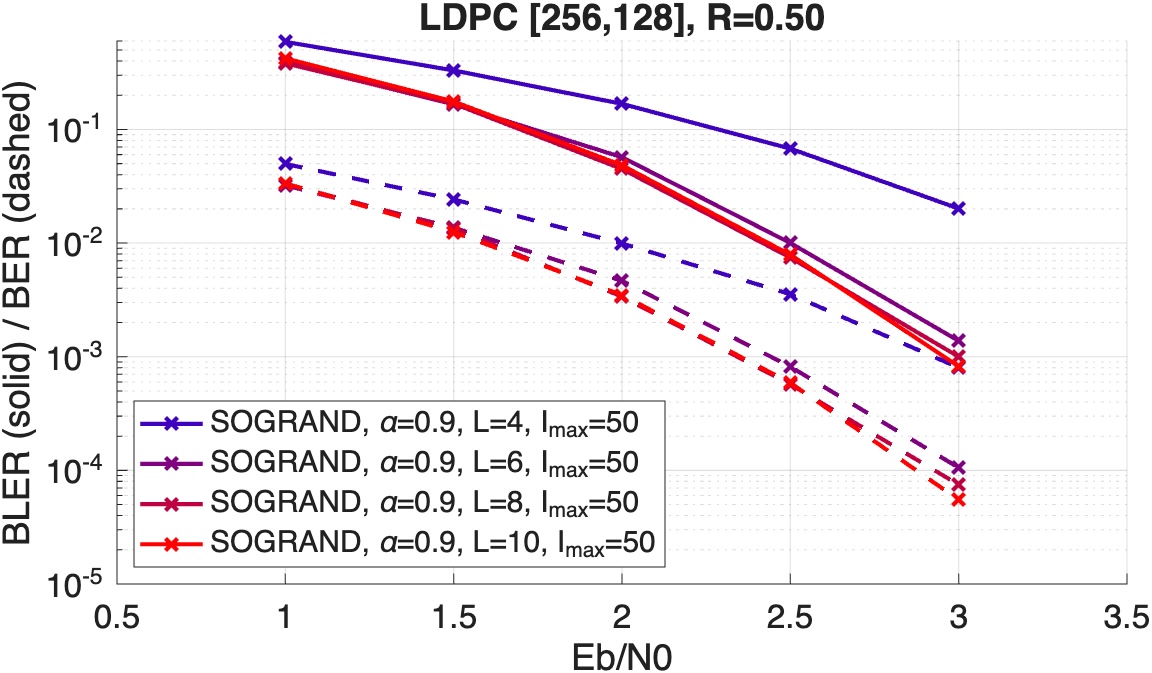}
\end{center}
\caption{The LDPC(256,128) rate $1/2$ code from 5G NR decoded with SOGRAND, $\alpha=0.9$, and a collection of list sizes, $L\in\{4,6,8,10\}$.} 
\label{fig:256 128 L}
\end{figure}

\begin{figure}[h]
\begin{center}
\includegraphics[width=1\columnwidth]{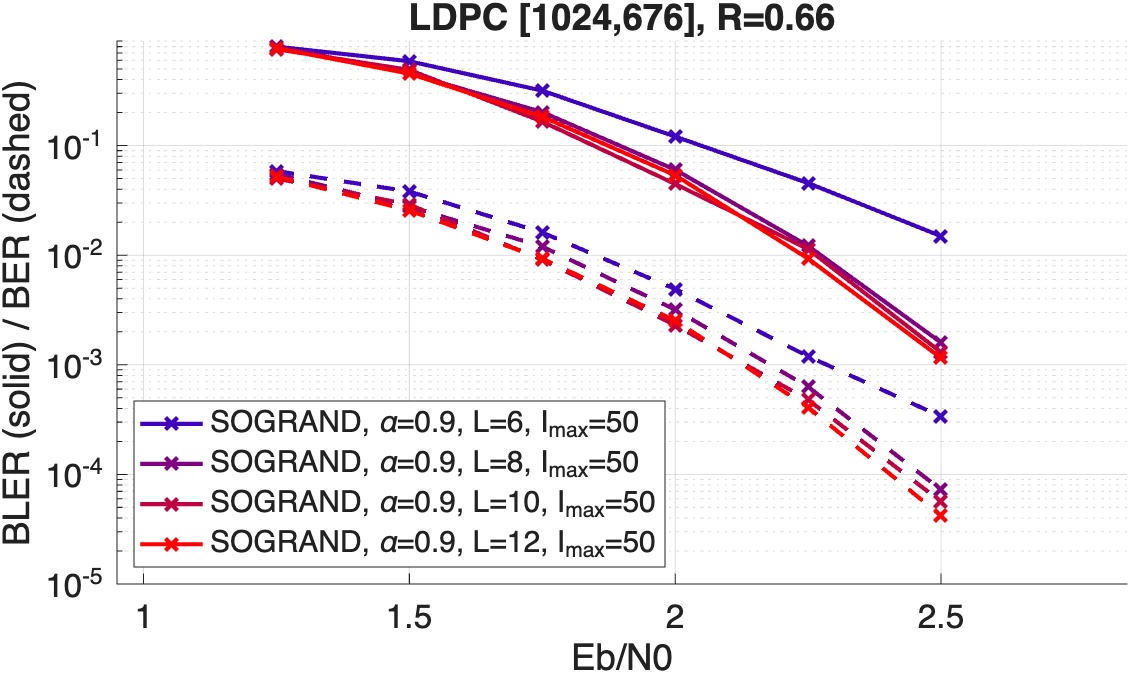}
\end{center}
\caption{The LDPC(1024,676) rate $2/3$ code from 5G NR decoded with SOGRAND, $\alpha=0.9$, and a collection of list sizes, $L\in\{6,8,10,12\}$.} 
\label{fig:1024 676 L}
\end{figure}

To assess the SOGRAND CN LDPC update we apply it with LDPC codes used in 5G NR in a BPSK channel subject to additive white Gaussian noise, allowing $I_{\max}=50$ iterations in all cases. As results for the two SOGRAND CN update rules, i.e. either using the even property or not, are essentially identical, here we show figures with the even CN rule, with the final figure showing results for both. For SOGRAND decoding we find that the scaling of extrinsic information has a minor impact, of order $0.05$ dB, for $\alpha$ selected between $0.8 - 1$ (data not shown), and so set $\alpha=0.9$ for the rest of the results without further optimization. For SPA and NMS, we use the implementations from MATLAB 5G NR toolbox.

We first consider the impact of SOGRAND's list size, $L$, on the block error rate (BLER) and bit error rate (BER) decoding performance. Fig. \ref{fig:256 128 L} and Fig. \ref{fig:1024 676 L} provide representative examples with codes of length $n=256$ and rate $1/2$, and $n=1024$ and rate $2/3$, respectively. In these figures, it can be seen that once the list size, $L$, is $8$ or larger, only minuscule improvement is observed with increasing list size.

\begin{figure}[h]
\begin{center}
\includegraphics[width=1\columnwidth]{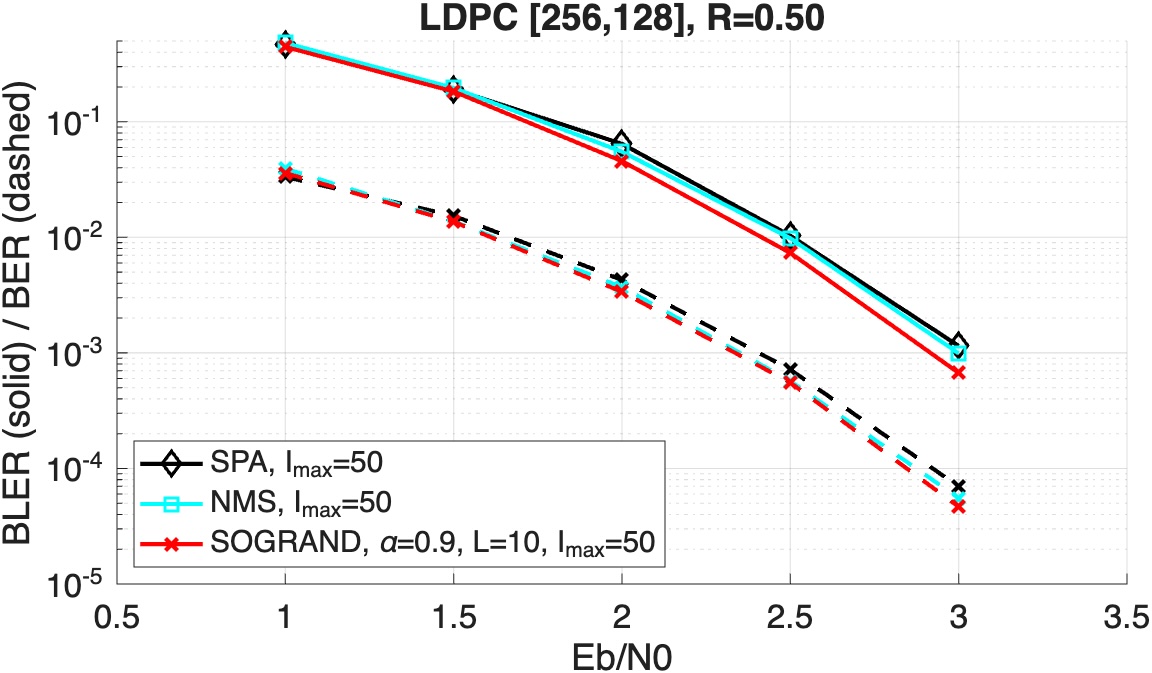}
\end{center}
\caption{The LDPC(256,128) rate $1/2$ code from 5G NR decoded with SPA, norm-min-sum, and SOGRAND with $\alpha=0.9$ and $L=10$.} 
\label{fig:256 128}
\end{figure}

\begin{figure}[h]
\begin{center}
\includegraphics[width=1\columnwidth]{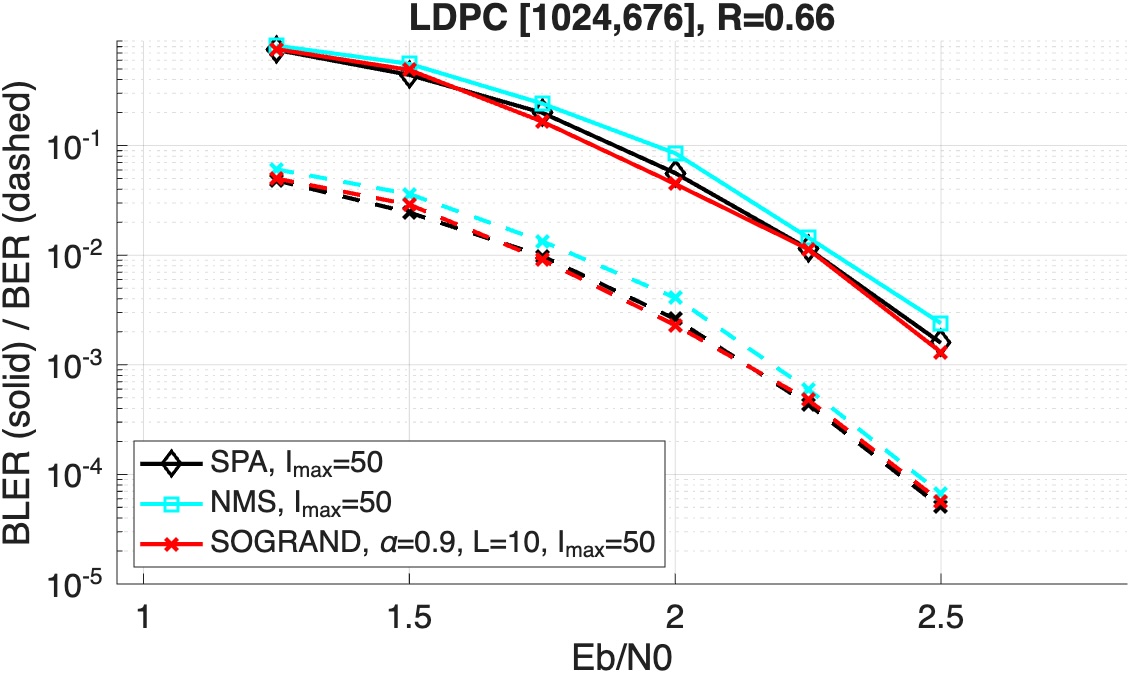}
\end{center}
\caption{The LDPC(1024,676) rate $2/3$ code from 5G NR decoded with SPA, norm-min-sum, and SOGRAND with $\alpha=0.9$ and $L=10$.} 
\label{fig:1024 676}
\end{figure}

We then establish the smallest list size necessary to meet or outperform the well established CN updates, SPA and norm-min-sum, where it can be seen that $L=10$ suffices, see Fig. \ref{fig:256 128} and Fig. \ref{fig:1024 676}. If basic ORBGRAND patterns were used, as in Fig. \ref{fig:SOGRAND seqs}, it would thus suffice to determine the rank ordered reliability of no more than the $8$ least reliable bits. Regardless of whether the even or non-even rule was used, a small table of $20$ patterns patterns suffices. If tables are not desired, published hardware \cite{Riaz25ORBGRAND,Kizilates26SOGRAND} shows it is possible to regenerate these sequences on the fly in a single clock cycle in a small circuit through parallelized implementation of the landslide algorithm \cite{duffy2022_ordered}.

\begin{figure}[h]
\begin{center}
\includegraphics[width=1\columnwidth]{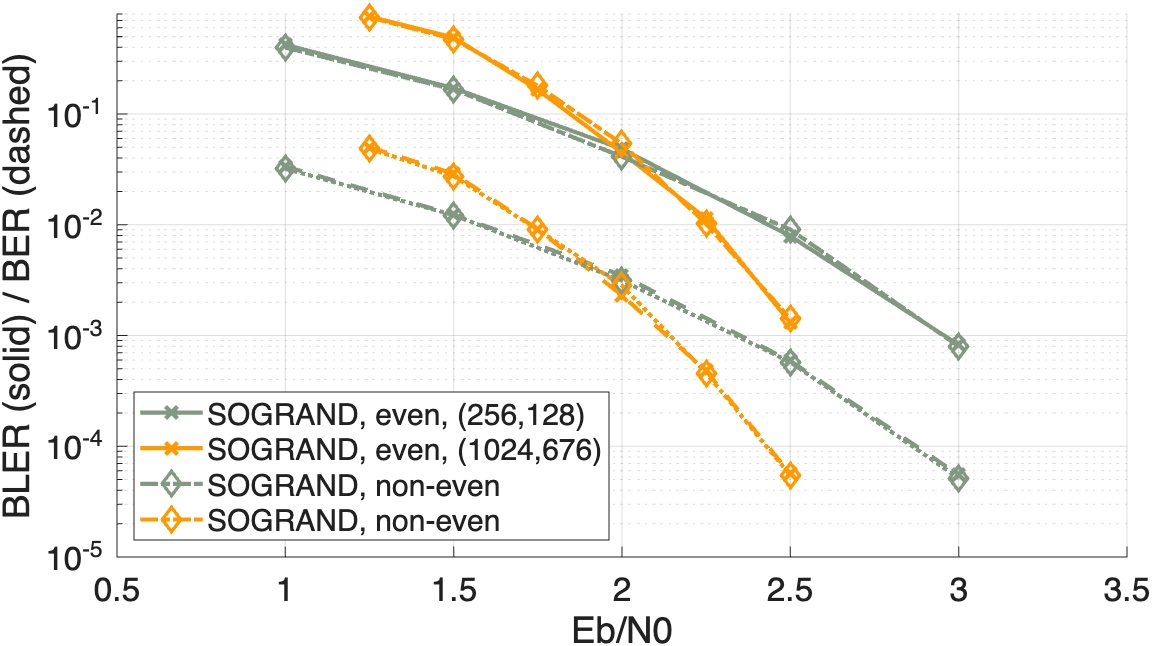}
\end{center}
\caption{The LDPC(256,128) rate $1/2$ and LDPC(1024,676) rate $2/3$ codes from 5G NR decoded with SOGRAND with $\alpha=0.9$, $L=10$, and each of the even and non-even CN update rules.} 
\label{fig:even}
\end{figure}
For the results presented so far, the CN update rule avails of the fact that an SPC code is even to perform only $L$ likelihood computations, but with the requirement that $\Psi$ in eq. \eqref{eq:Psi} be calculated. Alternatively, if the even nature of the SPC code is not used, one can circumvent calculating $\Psi$ by making $2L$ likelihood evaluations, i.e. both the even and odd parity queries, and obtain identical decoding performance, Fig. \ref{fig:even}. 

\section{Discussion}
\label{sec:discussion}
By specializing the recently introduced SOGRAND algorithm \cite{yuanSOGRAND25} to SPC codes, we have established two new means of evaluating a CN update for LDPC decoding that are hardware friendly and perform as well or better than existing methods. While this work establishes a new means of decoding LDPC codes by replacing the CN update, SOGRAND's ability to assess multiple parity checks simultaneously offers opportunities for additional developments.

\bibliographystyle{IEEEtran}
\bibliography{references}

@article{forney1968_exponential,
	title={Exponential error bounds for erasure, list, and decision feedback schemes},
	author={Forney, G},
	journal={{IEEE Trans. Inf. Theory}},
	volume={14},
	number={2},
	pages={206--220},
	year={1968},
	publisher={IEEE}
}

@article{gallagher1962_lowdensity,
  author={Gallager, R.},
  journal={IRE Trans. Inf. Theory}, 
  title={Low-density parity-check codes}, 
  year={1962},
  volume={8},
  pages={21-28},
}

@article{duffy_capacity-achieving_2019,
	title = {Capacity-Achieving Guessing Random Additive Noise Decoding},
	volume = {65},
	number = {7},
	fjournal = {IEEE Transactions on Information Theory},
	journal = {IEEE Trans. Inf. Theory},
	author = {Duffy, Ken R. and Li, Jiange and Medard, Muriel},
	year = {2019},
	pages={4023--4040}
}

@inproceedings{solomon20,
  title={Soft Maximum Likelihood Decoding using {GRAND}},
  author={A. Solomon and K. R. Duffy and M. M\'edard},
  booktitle={IEEE Int. Commun. Conf.},
  year={2020},
}

@inproceedings{riaz2021multicodegrand,
    author={Riaz, Arslan and Bansal, Vaibhav and Solomon, Amit and An, Wei and Liu, Qijun and Galligan, Kevin and Duffy, Ken R. and Medard, Muriel and Yazicigil, Rabia Tugce},
    booktitle={IEEE ESSCIRC},
    title={Multi-Code Multi-Rate Universal Maximum Likelihood Decoder using {GRAND}},
    year={2021},
}

@article{duffy2022_ordered,
  title={Ordered reliability bits guessing random additive noise decoding},
  author={Duffy, Ken R. and An, Wei and Medard, Muriel},
  journal={IEEE Trans. Signal Proc.},
  pages={4528 -- 4542},
  volume={70},
  year={2022},
}

@article{abbas2021orbgrand,
    author={Abbas, Syed Mohsin and Tonnellier, Thibaud and Ercan, Furkan and Jalaleddine, Marwan and Gross, Warren J.},
    journal={IEEE Trans. on VLSI Sys.},
    title={{High-Throughput and Energy-Efficient VLSI Architecture for Ordered Reliability Bits GRAND}},
    year={2022},
    volume={30},
    number={6}
}

@ARTICLE{Riaz25ORBGRAND,
  author={Riaz, Arslan and Yasar, Alperen and Ercan, Furkan and An, Wei and Ngo, Jonathan and Galligan, Kevin and M\'edard, Muriel and Duffy, Ken R. and Yazicigil, Rabia Tugce},
  fjournal={IEEE Journal of Solid-State Circuits}, 
  journal = {IEEE J. Solid-State Circuits},
  title={A Sub-0.8-{pJ}/bit universal Soft-Detection Decoder Using {ORBGRAND}}, 
  year={2025},
  volume={60},
  number={7},
  pages={2645--2659}
  }

@inproceedings{Kizilates25ORBRGANDAI,
  title={Low-latency modulation- and correlation-Adaptive {ORBGRAND-AI} decoder},
  author={E. Kizilates and A. Riaz and A. Bali and M. Grundei and M. M\'edard and K. R. Duffy and R. T. Yazicigil},
  booktitle={IEEE ESSERC},
  year={2025},
}

@article{condo2021fixed,
    author={Condo, Carlo},
    journal={IEEE Trans. Circuits Sys. I: Regular Papers},
    title={A Fixed Latency {ORBGRAND} Decoder Architecture With {LUT}-Aided Error-Pattern Scheduling},
    year={2022},
    volume={69},
    number={5},
    pages={2203--2211}
}

@article{rowshan2025segmented,
  title={Segmented {GRAND}: complexity Reduction through Sub-Pattern Combination},
  author={Rowshan, Mohammad and Yuan, Jinhong},
  fjournal={IEEE Transactions on Communications},
  journal={IEEE Trans. Commun.},
  year={2025},
  volume={73},
  number={8},
  pages={5607-5620},

}

@article{pyndiah_1998,
	title = {Near-optimum decoding of product codes: block turbo codes},
	volume = {46},
	number = {8},
	fjournal = {IEEE Transactions on Communications},
	journal = {IEEE Trans. Commun.},
	author = {Pyndiah, R.M.},
	year = {1998},
	pages={1003--1010},
}

@book{moon2020error,
  title={Error correction coding: mathematical methods and algorithms},
  author={Moon, Todd K},
  year={2020},
  publisher={John Wiley \& Sons}
}

@article{rowshan2024channel,
  title={Channel coding toward {6G}: Technical overview and outlook},
  author={Rowshan, Mohammad and Qiu, Min and Xie, Yixuan and Gu, Xinyi and Yuan, Jinhong},
  vjournal={IEEE Open Journal of the Communications Society},
  journal={IEEE Open J. Commun. Soc.},
  volume={5},
  pages={2585--2685},
  year={2024},
}

@book{richardson2008modern,
  title={Modern coding theory},
  author={Richardson, Tom and Urbanke, Ruediger},
  year={2008},
  publisher={Cambridge University Press}
}

@inproceedings{berrou_1993_turbo,
  author={Berrou, C. and Glavieux, A. and Thitimajshima, P.},
  booktitle={IEEE ICC}, 
  title={Near {Shannon} limit error-correcting coding and decoding: Turbo-codes}, 
  year={1993},
}

@article{smith2012staircase,
  author={B. P. {Smith} and A. {Farhood} and A. {Hunt} and F. R. {Kschischang} and J. {Lodge}},
  journal={J. Light. Technol.},
  fjournal={Journal of Lightwave Technology},
  title={Staircase {C}odes: {FEC} for 100 {G}b/s {OTN}}, 
  year={2012},
  volume={30},
  number={1},
  pages={110-117}
}

@ARTICLE{An22,
  author={An, Wei and M\'edard, Muriel and Duffy, Ken R.},
  journal={IEEE Trans. Commun.}, 
  title={Keep the bursts and ditch the interleavers}, 
  year={2022},
  volume={70},
  number={6},
  pages={3655--3667},
  }

@inproceedings{Duffy23ORBGRANDAI,
  author = {Duffy, Ken R. and Grundei, Moritz and M\'edard, Muriel},
  title = {Using channel correlation to improve decoding -- {ORBGRAND-AI}},
  year = {2023},
  booktitle = {IEEE Globecom}
}

@article{abbas2021list,
  title={{List-GRAND}: A practical way to achieve Maximum Likelihood Decoding},
  author={Abbas, Syed Mohsin and Jalaleddine, Marwan and Gross, Warren J},
  journal={IEEE Trans. Very Large Scale Integr. Syst.}, 
  number = {1},
  volumer = {31},
  pages = {43--54},
  year={2022}
}

@article{chu2023efficient,
  title={An efficient hard-detection GRAND decoder for systematic linear block codes},
  author={Chu, Shao-I and Ke, Syuan-An and Liu, Sheng-Jung and Lin, Yan-Wei},
  fjournal={IEEE Transactions on Very Large Scale Integration (VLSI) Systems},
  journal={IEEE Trans. Very Large Scale Integr. Syst.}, 
  volume={31},
  number={11},
  pages={1852--1864},
  year={2023},
}

@article{yuanSOGRAND25,
  author={Yuan, Peihong and M\'edard, Muriel and Galligan, Kevin and Duffy, Ken R.},
  fjournal={IEEE Transactions on Wireless Communications}, 
  journal={IEEE Trans. Wireless Commun.}, 
  title={Soft-output {(SO) GRAND} and Iterative Decoding to Outperform {LDPC} Codes}, 
  year={2025},
  volume={24},
  number={4},
  pages={3386--3399},
  }

@article{duffy2025soft,
  title={Soft-output guessing codeword decoding},
  author={Duffy, Ken R and Yuan, Peihong and Griffin, Joseph and M{\'e}dard, Muriel},
  fjournal={IEEE Communications Letters},
  journal={IEEE Commun. Lett.},
  year={2024},
  volume={29},
  number={2},
  pages={328--332},
  publisher={IEEE}
}

@article{yuan2025soft,
  title={Soft-Output Successive Cancellation List Decoding},
  author={Yuan, Peihong and Duffy, Ken R and M{\'e}dard, Muriel},
  fjournal={IEEE Transactions on Information Theory},
  journal={IEEE Trans. Inf. Theory},
  year={2025},
  volume={71},
  number={2},
  pages={1007--1017},
}

@inproceedings{blanc2024GRANDABDecoder,
  title = {A {GRANDAB} decoder with 8.48 {Gbps} worst-case throughput in 65nm {{CMOS}}},
  booktitle = {IEEE ESSERC},
  author = {Blanc, Ludovic D. and Herrmann, Victor and Ren, Yuqing and M{\"u}ller, Christoph and Kristensen, Andreas T. and Levisse, Alexandre and Shen, Yifei and Burg, Andreas},
  year = {2024},
  month = sep,
  pages = {685--688},
}

@article{feng2025SO,
  title={Leveraging Code Structure to Improve Soft Output for {GRAND, GCD, OSD, and SCL}},
  author={Feng, Jiewei and Duffy, Ken R and M{\'e}dard, Muriel },
  journal={arXiv:2503.16677},
  year={2025}
}

@article{ji2025efficient,
  title={Efficient {ORBGRAND} implementation with parallel noise sequence generation},
  author={Ji, Chao and You, Xiaohu and Zhang, Chuan and Studer, Christoph},
  fjournal={IEEE Transactions on Very Large Scale Integration (VLSI) Systems},
  journal={IEEE Trans. Very Large Scale Integr. (VLSI) Syst.},
  year={2025},
  volume={33},
  number={2},
  pages={435-448},
}

@article{costello2007channel,
  title={Channel coding: The road to channel capacity},
  author={Costello, Daniel J and Forney, G David},
  fjournal={Proceedings of the IEEE},
  journal={Proc. IEEE},
  volume={95},
  number={6},
  pages={1150--1177},
  year={2007},
}

@article{mackay1997near,
  title={Near {S}hannon limit performance of low density parity check codes},
  author={MacKay, David JC and Neal, Radford M},
  fjournal={Electronics letters},
  journal = {Electron. Lett.},
  volume={33},
  number={6},
  pages={457--458},
  year={1997}
}

@article{richardson2001efficient,
  title={Efficient encoding of low-density parity-check codes},
  author={Richardson, Thomas J and Urbanke, R{\"u}diger L},
  fjournal={IEEE transactions on information theory},
    journal={IEEE Trans. Inf. Theory},
  volume={47},
  number={2},
  pages={638--656},
  year={2001},
  publisher={IEEE}
}

@article{mansour2003high,
  title={High-throughput {LDPC} decoders},
  author={Mansour, Mohammad M and Shanbhag, Naresh R},
  fjournal={IEEE transactions on very large scale integration (VLSI) Systems},
  journal={IEEE Trans. Very Large Scale Integr. VLSI Syst.},
  volume={11},
  number={6},
  pages={976--996},
  year={2003},
}

@article{hailes2015survey,
  title={A survey of {FPGA}-based {LDPC} decoders},
  author={Hailes, Peter and Xu, Lei and Maunder, Robert G and Al-Hashimi, Bashir M and Hanzo, Lajos},
  fjournal={IEEE Communications Surveys \& Tutorials},
  journal={IEEE Commun. Surv. Tutor.},
  volume={18},
  number={2},
  pages={1098--1122},
  year={2015},
  publisher={IEEE}
}

@article{richardson2001capacity,
  title={The capacity of low-density parity-check codes under message-passing decoding},
  author={Richardson, Thomas J and Urbanke, R{\"u}diger L},
  journal={IEEE Trans. Inf. Theory},
  volume={47},
  number={2},
  pages={599--618},
  year={2001},
}

@article{zhang2010efficient,
  title={An efficient {10GBASE-T} ethernet {LDPC} decoder design with low error floors},
  author={Zhang, Zhengya and Anantharam, Venkat and Wainwright, Martin J and Nikolic, Borivoje},
  fjournal={IEEE Journal of Solid-State Circuits},
  journal={IEEE J. of Solid-State Circuits},
  volume={45},
  number={4},
  pages={843--855},
  year={2010},
}

@book{mackay2003information,
  title={Information theory, inference and learning algorithms},
  author={MacKay, David JC},
  year={2003},
  publisher={Cambridge university press}
}

@article{richardson2018design,
  title={Design of low-density parity check codes for {5G} new radio},
  author={Richardson, Tom and Kudekar, Shrinivas},
  fjournal={IEEE Communications Magazine},
  journal={IEEE Commun. Mag.},
  volume={56},
  number={3},
  pages={28--34},
  year={2018},
}

@article{abbas2025improved,
  title={Improved Step-{GRAND}: low-latency soft-input guessing random additive noise decoding},
  author={Abbas, Syed Mohsin and Jalaleddine, Marwan and Tsui, Chi-Ying and Gross, Warren J},
  fjournal={IEEE Transactions on Very Large Scale Integration (VLSI) Systems},
  journal={IEEE Trans. Very Large Scale Integr. (VLSI) Syst.},
  year={2025},
  volume={33},
  number={4},
  pages={1028--1041},
}

@ARTICLE{chatzigeorgiou2023symbol,
  author={Chatzigeorgiou, Ioannis and Monteiro, Francisco A.},
  fjournal={IEEE Communications Letters}, 
  journal={IEEE Commun. Lett.}, 
  title={Symbol-Level {GRAND} for High-Order Modulation Over Block Fading Channels}, 
  year={2023},
  volume={27},
  number={2},
  pages={447--451},
}

@article{chatzigeorgiou2024guessing,
  title={Guessing random additive noise decoding of network coded data transmitted over burst error channels},
  author={Chatzigeorgiou, Ioannis and Savostyanov, Dmitry},
  fjournal={IEEE Transactions on Vehicular Technology},
  journal={ IEEE Trans. Veh. Technol.},
  volume={73},
  number={9},
  pages={12842--12857},
  year={2024},
}

@article{ma2024guessing,
  title={Guessing what, noise or codeword?},
  author={Ma, Xiao},
  journal={IEEE ITW},
  year={2024}
}

@INPROCEEDINGS{Janz25soft,
  author={Janz, Tim and Obermüller, Simon and Zunker, Andreas and Ten Brink, Stephan},
  fbooktitle={International Symposium on Topics in Coding (ISTC)}, 
  booktitle={ISTC}, 
  title={Soft-Output from covered space decoding of product codes}, 
  year={2025},
}

@inproceedings{Kizilates26SOGRAND,
  title={Soft output threshold-guided {CRC} decoding with {SOGRAND} in 40nm {CMOS}},
  author={E. Kizilates and A. Riaz and A. Bali and J. Feng and M. M\'edard and K. R. Duffy and R. T. Yazicigil},
  booktitle={IEEE CICC},
  year={2026},
}

@article{duffy2026guessing,
  title={Guessing random additive noise decoding for digital data communication},
  author={Duffy, Ken R and M{\'e}dard, Muriel},
  fjournal={Foundations and Trends{\textregistered} in Integrated Circuits and Systems},
  journal={Found. Trends Integr. Circuits Syst.},
  volume={5},
  number={2},
  pages={105--215},
  year={2026},
  publisher={Emerald Publishing Limited}
}

@ARTICLE{renAGeneralized2024,
  author={Ren, Yuqing and Harb, Hassan and Shen, Yifei and Balatsoukas-Stimming, Alexios and Burg, Andreas},
  fjournal={IEEE Transactions on Circuits and Systems I: Regular Papers}, 
  journal={IEEE TCAS-I},
  title={A Generalized Adjusted Min-Sum Decoder for {5G LDPC} Codes: Algorithm and Implementation}, 
  year={2024},
  volume={71},
  number={6},
  pages={2911-2924},
}

@article{jang2024area,
  title={Area-efficient {QC-LDPC} decoding architecture with thermometer code-based sorting and relative quasi-cyclic shifting},
  author={Jang, Boseon and Jang, Hyejung and Kim, Sungho and Choi, Kangjoon and Park, In-Cheol},
  fjournal={IEEE Transactions on Circuits and Systems I: Regular Papers},
  journal={IEEE TCAS-I},
  volume={71},
  number={6},
  pages={2897--2910},
  year={2024},
}

@article{lee2022multi,
  title={Multi-mode QC-LDPC decoding architecture with novel memory access scheduling for 5G new-radio standard},
  author={Lee, Seongjin and Park, Sangsoo and Jang, Boseon and Park, In-Cheol},
  fjournal={IEEE Transactions on Circuits and Systems I: Regular Papers},
  journal={IEEE TCAS-I},
  volume={69},
  number={5},
  pages={2035--2048},
  year={2022},
  publisher={IEEE}
}
\end{document}